\documentclass[aps, prl, twocolumn, a4paper, reprint, floatfix]{revtex4-1}

\usepackage[utf8]{inputenc}
\usepackage{epsfig}
\usepackage[T1]{fontenc}
\usepackage{t1enc}
\usepackage{graphicx}
\usepackage{amssymb}
\usepackage{amsmath}
\usepackage{relsize}
\usepackage{overpic}
\usepackage{bm}
\usepackage{hyperref}

\usepackage[normalem]{ulem}
\usepackage{pdfpages}

\newcommand{\avg}[1]{{\left<#1\right>}}

\usepackage{mathtools}
\def\multiset#1#2{\ensuremath{\left(\kern-.3em\left(\genfrac{}{}{0pt}{}{#1}{#2}\right)\kern-.3em\right)}}

\usepackage{amsmath}
\DeclareMathOperator{\Tr}{Tr}

\begin{document}

\title{Eigenvalue Spectra of Modular Networks}

\author{Tiago P. Peixoto}
\email{tiago@itp.uni-bremen.de}
\affiliation{Institut f\"{u}r Theoretische Physik, Universit\"{a}t Bremen, Hochschulring 18, D-28359 Bremen, Germany}


\begin{abstract}
A large variety of dynamical processes that take place on networks can
be expressed in terms of the spectral properties of some linear operator
which reflects how the dynamical rules depend on the network
topology. Often such spectral features are theoretically obtained by
considering only local node properties, such as degree
distributions. Many networks, however, possess large-scale modular
structures that can drastically influence their spectral
characteristics, and which are neglected in such simplified
descriptions. Here we obtain in a unified fashion the spectrum of a
large family of operators, including the adjacency, Laplacian and
normalized Laplacian matrices, for networks with generic modular
structure, in the limit of large degrees. We focus on the conditions
necessary for the merging of the isolated eigenvalues with the
continuous band of the spectrum, after which the planted modular
structure can no longer be easily detected by spectral methods. This is
a crucial transition point which determines when a modular structure is
strong enough to affect a given dynamical process. We show that this
transition happens in general at different points for the different
matrices, and hence the detectability threshold can vary significantly
depending on the operator chosen. Equivalently, the sensitivity to the
modular structure of the different dynamical processes associated with
each matrix will be different, given the same large-scale structure
present in the network. Furthermore, we show that, with the exception of
the Laplacian matrix, the different transitions coalesce into the same
point for the special case where the modules are homogeneous, but
separate otherwise.
\end{abstract}

\maketitle

Networks form the substrate of a dominating class of interacting complex
systems, on which various dynamical processes take place. Many of the
most important types of dynamics such as random
walks~\cite{noh_random_2004, samukhin_laplacian_2008}, diffusion,
synchronization~\cite{barahona_synchronization_2002,arenas_synchronization_2008,
almendral_dynamical_2007} and epidemic
spreading~\cite{wang_epidemic_2003,
castellano_thresholds_2010,goltsev_localization_2012} have central
properties which are directly expressed via the spectral features of
matrices associated with the network
topology~\cite{dorogovtsev_spectra_2003,chung_spectra_2003,kim_ensemble_2007},
such as the mixing time of random walks, epidemic thresholds and the
synchronization speed of oscillators, to name a few. Virtually all of
these processes will be affected by large-scale modular structures
present in the network~\cite{newman_communities_2011}, which is
reflected in its spectral properties~\cite{ergun_spectra_2009,
kuhn_spectra_2011, chauhan_spectral_2009, nadakuditi_graph_2012}. Since
such large-scale modularity is a ubiquitous property in real
networks~\cite{newman_communities_2011}, describing the spectral
features resulting from this is a crucial step in understanding how
these systems function. Additionally, the information encoded in the
eigenvectors of these matrices are central to the nontrivial task of
detecting large-scale features in empirical
networks~\cite{fortunato_community_2010, newman_finding_2006,
fiedler_algebraic_1973, pothen_partitioning_1990,
nadakuditi_graph_2012}, and from it is possible to derive general bounds
on the detectability of existing community
structure~\cite{nadakuditi_graph_2012}.

In this work, we formulate an unified framework to obtain the eigenvalue
spectrum associated with arbitrary modular structures, parameterized as
stochastic block models~\cite{holland_stochastic_1983,
fienberg_statistical_1985, faust_blockmodels:_1992,
karrer_stochastic_2011}. The framework allows the straightforward
calculation of a large class of matrices which include the adjacency,
Laplacian and normalized Laplacian matrices, and is exact in the limit
of large degrees. It contrasts with previous
work~\cite{kuhn_spectra_2011} which is exact in the limit of small
degrees, but depends on the solution of a number of self-consistency
equations which are solved stochastically. Here we show that if the
block structure is sufficiently well pronounced, it will trigger the
appearance of isolated eigenvalues, with associated eigenvectors
strongly correlated with the block partition. If the block structure
becomes too weak (but nonvanishing), the isolated eigenvalues merge
with the continuous band, and the eigenvectors are no longer correlated
with the block partition. This has important consequences to the
detectability of modular structure in
networks~\cite{nadakuditi_graph_2012} but also to a large class of
dynamical processes since after this transition takes place one should
not expect the modular structure to play a significant role. We show
that in general the different matrices have different sensitivities to
the imposed block structure, and exhibit these transitions for different
modularity strengths.

\begin{figure}
  \begin{overpic}[unit=1cm, width=1\columnwidth, trim=0.0cm 0 0 0, clip]{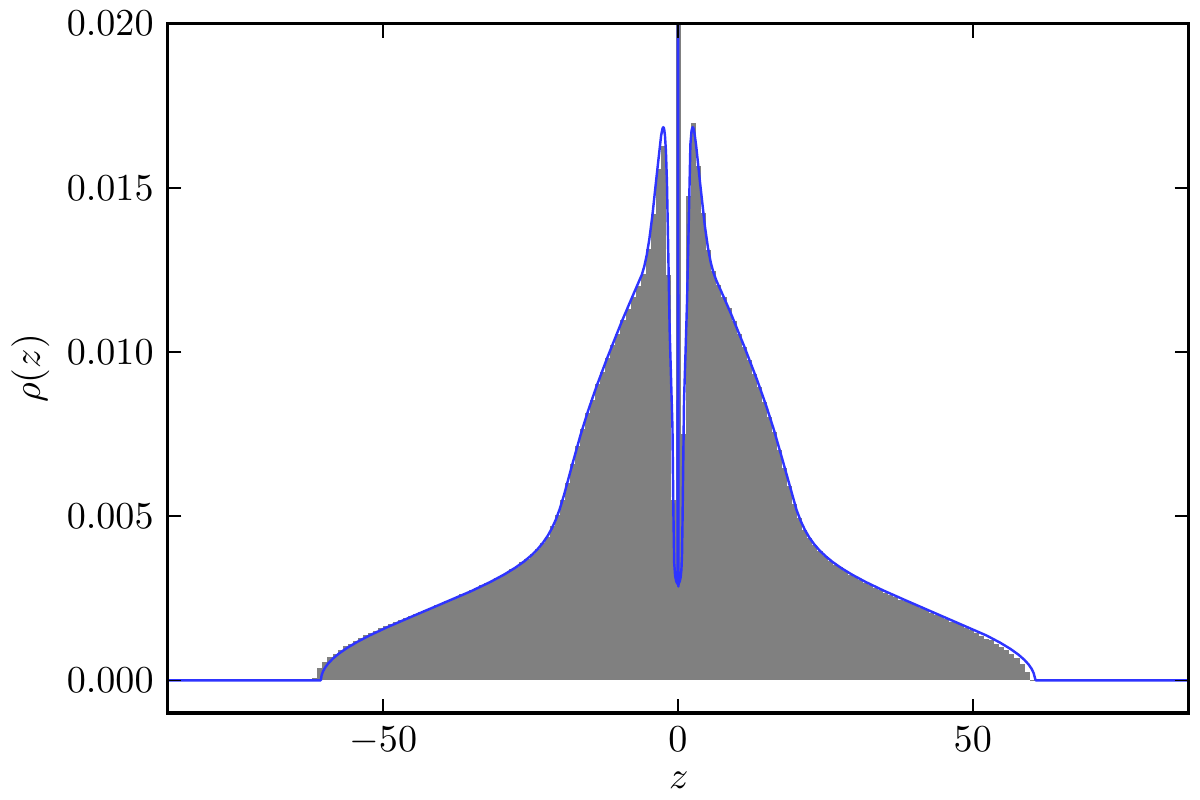}
    \put(18,32){\includegraphics[width=.29\columnwidth]{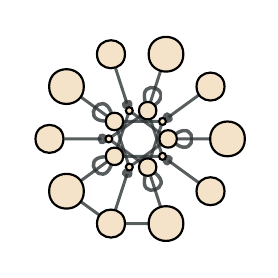}}
    \put(70,25){\includegraphics[width=.29\columnwidth]{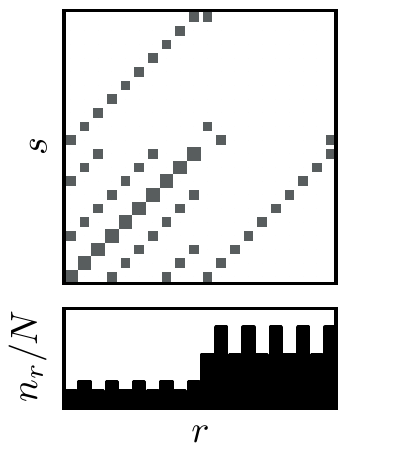}}
  \end{overpic}
  \includegraphics[width=1\columnwidth]{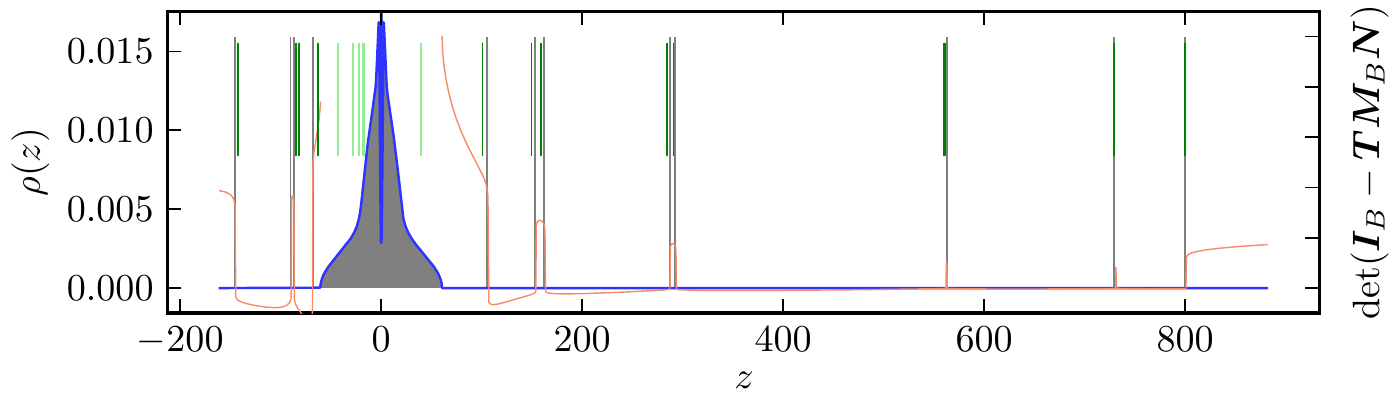} \caption{ \label{fig:example_adj}
  \emph{Top:} Continuous band of the matrix $\bm{A}$ for the block
  structure in the inset (right: $e_{rs}$ matrix and block sizes $n_r$,
  left: graphical representation). The solid line corresponds to
  Eq.~\ref{eq:rho}, and the grey histogram is averaged over $25$ network
  realizations with $N=2\times10^4$, and $\avg{k}=300$.\emph{Bottom:}
  The same, but with the isolated eigenvalues added. The grey vertical
  lines are average empirical values, whereas the solid (orange) curve
  corresponds to the determinant of Eq.~\ref{eq:detached}. The vertical
  (green) line segments mark the eigenvalues of the matrix $\bm{C}_B +
  \bm{M}_B\bm{N}$.}
\end{figure}

\begin{figure}
  \includegraphics[width=1\columnwidth]{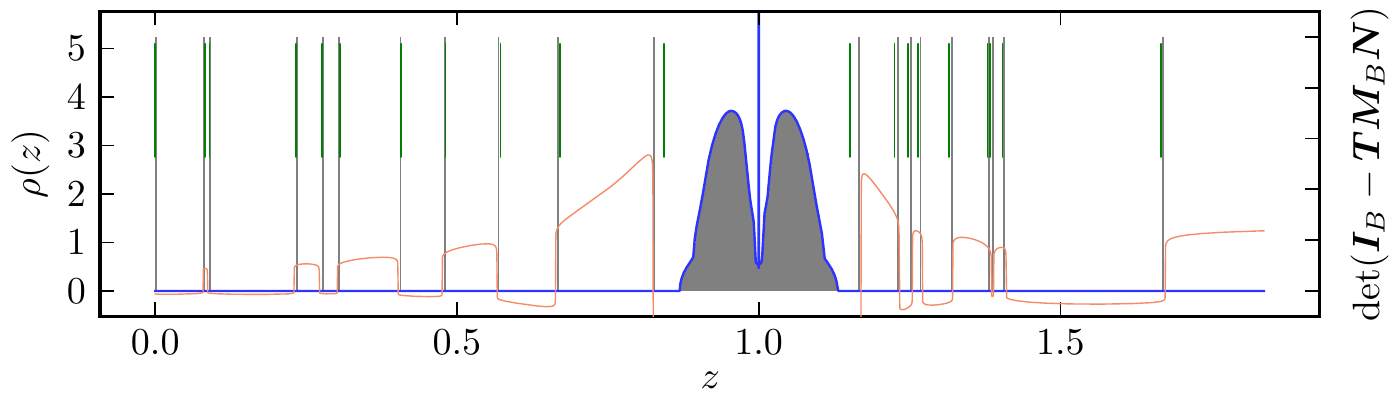}
  \includegraphics[width=1\columnwidth]{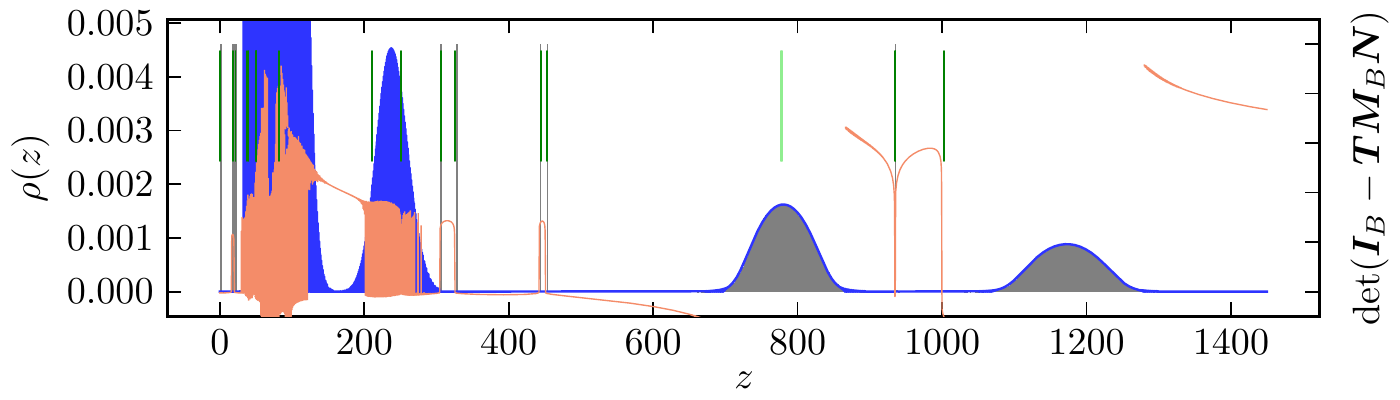} \caption{ \label{fig:example_nlap}Eigenvalue
  spectrum of the normalized Laplacian matrix $\bm{\mathcal{L}}$ (top)
  and Laplacian matrix $\bm{L}$ (bottom) for the block structure of
  Fig.~\ref{fig:example_adj}.}
\end{figure}

\emph{Unified framework. ---} Any given undirected network can be
encoded via its adjacency matrix $\bm{A}$, which has entries $A_{ij} =
1$ if node $i$ is adjacent to $i$, or $A_{ij} = 0$ otherwise. The
Laplacian matrix is defined as $\bm{L} = \bm{D} - \bm{A}$, where
$\bm{D}$ is a diagonal matrix containing the vertex degrees, $D_{ij} =
\delta_{ij}k_i$. Finally, the normalized Laplacian is defined as
$\bm{\mathcal{L}} = \bm{I} - \bm{D}^{-1/2}\bm{A}\bm{D}^{-1/2}$. Here we
use a general parametrization which contains these matrices as special
cases, via the matrix $\bm{W} = \bm{C} + \bm{M}$, where $\bm{C}$ is a
random diagonal matrix, and $\bm{M}$ is a random symmetric
matrix. Simply by choosing $\{\bm{C}=0, \bm{M}=\bm{A}\}$, $\{\bm{C}=D,
\bm{M}=-\bm{A}\}$ and $\{\bm{C}=\bm{I},
\bm{M}=-\bm{D}^{-1/2}\bm{A}\bm{D}^{-1/2}\}$, we recover $\bm{A}$,
$\bm{L}$ and $\bm{\mathcal{L}}$, respectively. We may write $\bm{W} =
\bm{C} + \bm{\mathcal{M}} + \avg{\bm{M}} = \bm{\mathcal{X}} +
\avg{\bm{M}}$, such that the matrix $\bm{\mathcal{X}} = \bm{C} +
\bm{\mathcal{M}}$, with $\bm{\mathcal{M}} = \bm{M} - \avg{\bm{M}}$, has
off-diagonal entries with zero mean. The spectrum of $\bm{\mathcal{X}}$
can be obtained via its average resolvent $\avg{(z\bm{I}
- \bm{\mathcal{X}})^{-1}}$, using the Stieltjes transform $\rho(z) =
  -\frac{1}{N\pi}\operatorname{Im}\Tr\avg{(z\bf{I} -
 \bm{\mathcal{X}})^{-1}}$, with $z$ approaching the real line from
above. Given an arbitrary random matrix $\bm{X}$ with zero-mean
off-diagonal entries, if the variance of the entries is sufficiently
large, we can use the approximation~\cite{nadakuditi_spectra_2013},
\begin{equation}\label{eq:inv}
  \avg{[\bm{X}^{-1}]_{ii}} \simeq \sum_{X_{ii}}\frac{P^i(X_{ii})}{X_{ii}
  - \sum_j\avg{[\bm{X}^{-1}]_{jj}}\avg{a^2_j}},
\end{equation}
and $\avg{[\bm{X}^{-1}]_{ij}} = 0$ for $i\neq j$, where $\bm{a}$ is the
$i$th column of $\bm{X}$, with the diagonal element removed, and it is
assumed that the diagonal elements $X_{ii}$ can only take discrete
values, distributed according to $P^i(X_{ii})$. We use Eq.~\ref{eq:inv}
to compute the average resolvent of the matrix $\bm{\mathcal{X}}$. We
consider random graphs parameterized as stochastic
block models~\cite{holland_stochastic_1983, fienberg_statistical_1985,
faust_blockmodels:_1992} where $N$ nodes are divided into $B$ distinct
blocks, where each block $r$ has $n_r$ nodes, and the matrix entry
$e_{rs}$ specifies the number of edges between blocks $r$ and $s$,
which are otherwise randomly placed. Hence, in the considered cases, the
expected value of $\bm{M}$ is simply a function of the block
memberships,
i.e. $\avg{C_{ii}} = [\bm{C}_B]_{b_i,b_i}=c_{b_i}$ and $\avg{M_{ij}} =
[\bm{M}_B]_{b_i, b_j}$, with $\bm{C}_B$ and $\bm{M}_B$ being matrices of
size $B\times B$, and the vector $\bm{b}$ of size $N$ and entries in the
range $[1,B]$ specifies the block memberships. When applying this to
Eq.~\ref{eq:inv} with $\bm{X} = z\bm{I}- \bm{\mathcal{X}}$, we may use
the fact the averages on both sides of Eq.~\ref{eq:inv} can only depend
on the block membership of the respective nodes. Thus, using the
shorthand $t_r(z) \equiv \avg{[(z\bm{I}- \bm{\mathcal{X}})^{-1}]_{ii}}$
for $i\in r$, we obtain,
\begin{equation}\label{eq:t}
  t_r(z) = \sum_c \frac{p^r_c}{z - c - \sum_s\sigma^2_{rs}n_st_s(z)},
\end{equation}
where $p^r_c$ is probability distribution of the diagonal elements $c$
for block $r$, and $\sigma^2_{rs}$ is the variance of the elements of
$\bm{\mathcal{M}}$, labeled according to block membership, which is
identical to the variance of $\bm{M}$. The spectrum of
$\bm{\mathcal{X}}$ may be finally obtained via
\begin{equation}\label{eq:rho}
  \rho(z) = -\frac{1}{N\pi}\sum_rn_r\operatorname{Im}t_r(z).
\end{equation}
In order to obtain the spectrum of $\bm{W}$, we employ an argument
developed in Ref.~\cite{benaych-georges_eigenvalues_2011}, and note that in
order for $z$ to be an eigenvalue of $\bm{W}=\bm{\mathcal{X}} + \bm{M}$,
we must have $\det(z\bm{I} - (\bm{\mathcal{X}} + \avg{\bm{M}})) = 0$,
which can be rewritten as $\det(z\bm{I} - \bm{\mathcal{X}})\det(\bm{I} -
(z\bm{I} - \bm{\mathcal{X}})^{-1}\avg{\bm{M}}) = 0$.  Thus, if the
second determinant is zero for a given $z$, it will be an eigenvalue of
$\bm{W}$ but not of $\bm{\mathcal{X}}$. These additional eigenvalues may
be obtained via the ensemble average $\det(\bm{I} - \avg{(z\bm{I} -
\bm{\mathcal{X}})^{-1}}\avg{\bm{M}}) = 0$, which will hold if the matrix
$\avg{(z\bm{I} - \bm{\mathcal{X}})^{-1}}\avg{\bm{M}}$ has an eigenvalue
equal to one. Since this matrix has a maximum rank equal to $B$, its
nonzero eigenvalues will be identical to the $B\times B$ matrix
$\bm{T}(z)\bm{M}_B\bm{N}$, where $\bm{T}(z)$ and $\bm{N}$ are diagonal
$B\times B$ matrices containing the values of $t_r(z)$ and $n_r$,
respectively. Hence, the existence of additional eigenvalues of $\bm{W}$
may obtained by solving,
\begin{equation}\label{eq:detached}
  \det(\bm{I}_B - \bm{T}(z)\bm{M}_B\bm{N}) = 0,
\end{equation}
simultaneously with $\rho(z) = 0$. Eqs.~\ref{eq:t}, \ref{eq:rho}
and~\ref{eq:detached} provide a complete recipe for obtaining the
desired spectrum, provided we know the $B\times B$ matrices
$\sigma^2_{rs}$ and $\bm{M}_B$ as well as the diagonal entry
distribution $p^r_c$. For the three matrices of interest they are easily
computed as $\{p^r_c = \delta_{0,c};\; \sigma^2_{rs} = [\bm{M}_B]_{rs} =
e_{rs} / n_rn_s\}$ for $\bm{A}$, $\{p^r_c = P(c, e_r/n_r);\;
[\bm{M}_B]_{rs} = -e_{rs} / n_rn_s;\; \sigma^2_{rs} = e_{rs} / n_rn_s\}$
for $\bm{L}$, with $P(c, \lambda)$ being a Poisson distribution on $c$
with average $\lambda$, and $\{p^r_c = \delta_{1,c};\; [\bm{M}_B]_{rs} =
-e_{rs}/ \sqrt{n_re_rn_se_s};\; \sigma^2_{rs} \simeq e_{rs} / e_re_s\}$
for $\bm{\mathcal{L}}$. We emphasize that, since the approximation in
Eq.~\ref{eq:inv} was used, the obtained spectrum should be correct only
in the limit of sufficiently large degrees. If this holds, the theory
reproduces in very good detail the spectrum of empirical networks, as
can be seen in Fig.~\ref{fig:example_adj}. The spectrum is composed of a
continuous band, as well as a number of isolated eigenvalues, which
correspond very well to the solutions of Eqs.~\ref{eq:rho}
and~\ref{eq:detached}, respectively. The same is true for the spectrum
of the matrices $\bm{\mathcal{L}}$ and $\bm{L}$
(Fig.~\ref{fig:example_nlap}). The spectrum of $\bm{L}$ is special,
since it contains an elaborate fine structure, with many fringes, and an
interleaving of the continuous band (Eq.~\ref{eq:rho}) with the isolated
eigenvalues (Eq.~\ref{eq:detached}). The continuous band has no
well-defined edge, with fringes which extend through the whole spectrum,
but with decaying amplitudes. Despite such detailed structure, the
theory captures these features very well, as can be seen in
Fig.~\ref{fig:example_nlap} (see also the Supplemental Material).

For isolated eigenvalues which are sufficiently detached from the
spectral band, Eq.~\ref{eq:t} may be approximated by $t_r \approx 1/(z -
c_r)$, in which case Eq.~\ref{eq:detached} amounts to $\det(z\bm{I}_B -
(\bm{C}_B + \bm{M}_B\bm{N})) = 0$, where $\bm{C}_B$ is a diagonal matrix
with the $c_r$ values. If this holds, the detached eigenvalues will
correspond to the spectrum of the matrix $\bm{C}_B + \bm{M}_B\bm{N}$.

At the edges of the continuous band the purely real solution to
Eq.~\ref{eq:t} becomes unstable, and the largest eigenvalue of the
Jacobian $J_{rs}(z)\equiv\partial \hat{t}_r/\partial t_s = \sum_c
p^r_c\sigma^2_{rs}n_s / (z - c - \sum_s\sigma^2_{rt}n_tt_t(z))^2$, where
$\hat{t}_r$ is the right-hand side of Eq.~\ref{eq:t}, becomes equal to
one. Hence, one may find the edges of the continuous band by solving
$\det(\bm{I}_B - \bm{J}(z)) = 0$, simultaneously with $\rho(z) = 0$.

\begin{figure}
  \includegraphics[width=.58\columnwidth]{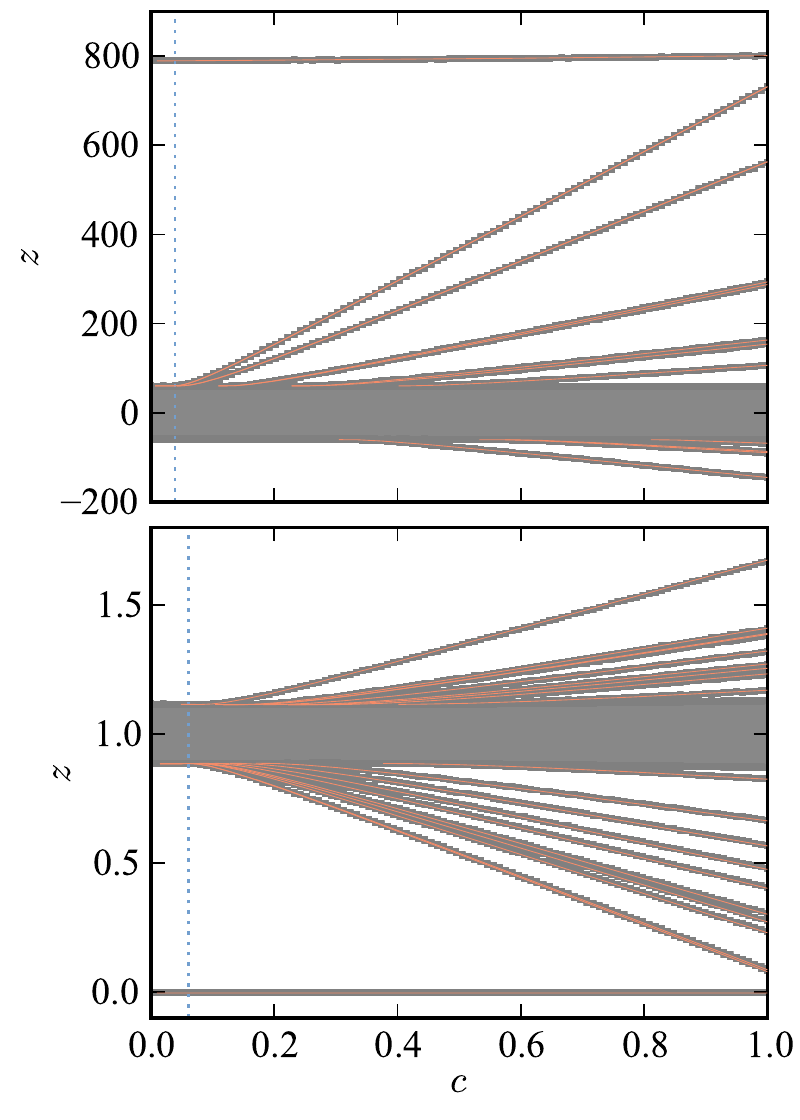}
  \begin{minipage}[b]{.4\columnwidth}
    \includegraphics[width=\columnwidth]{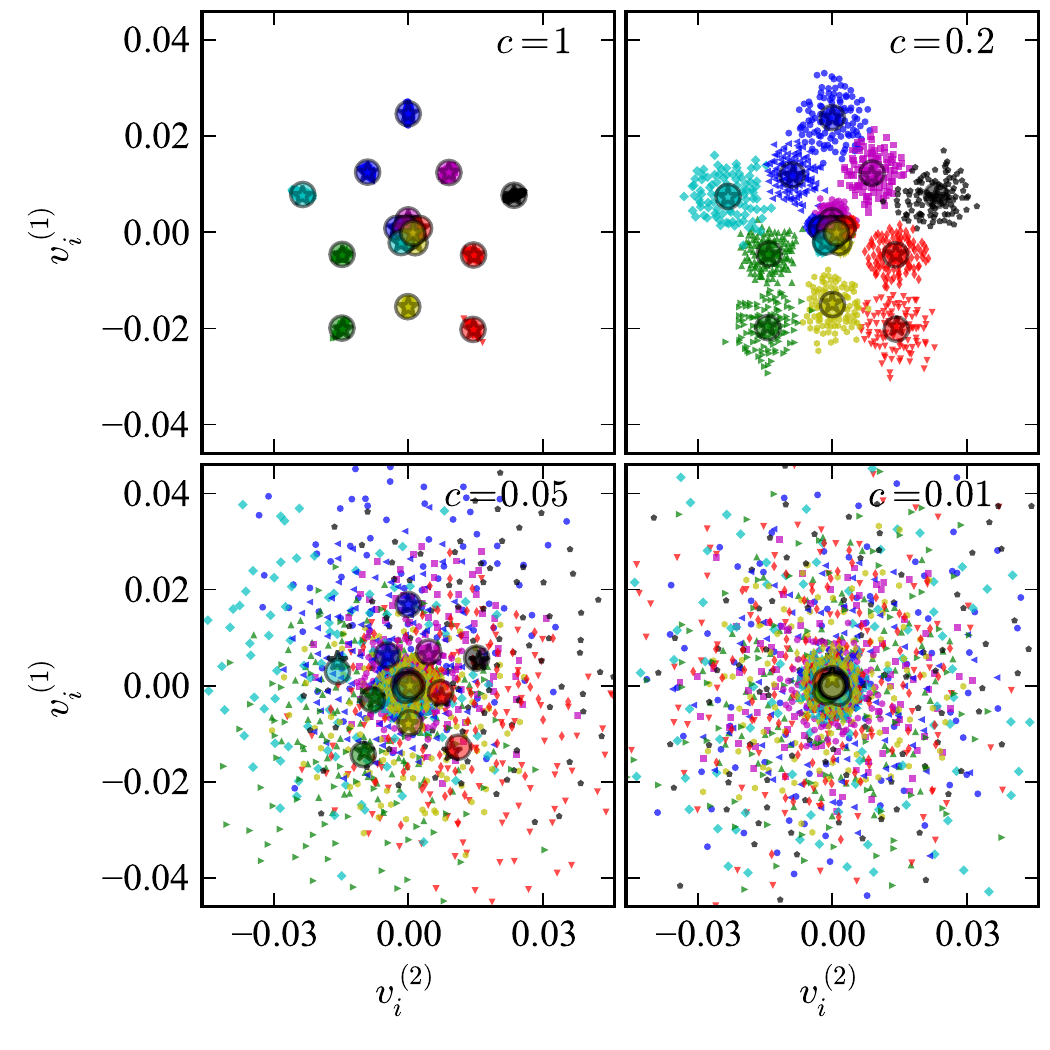}
    \includegraphics[width=\columnwidth]{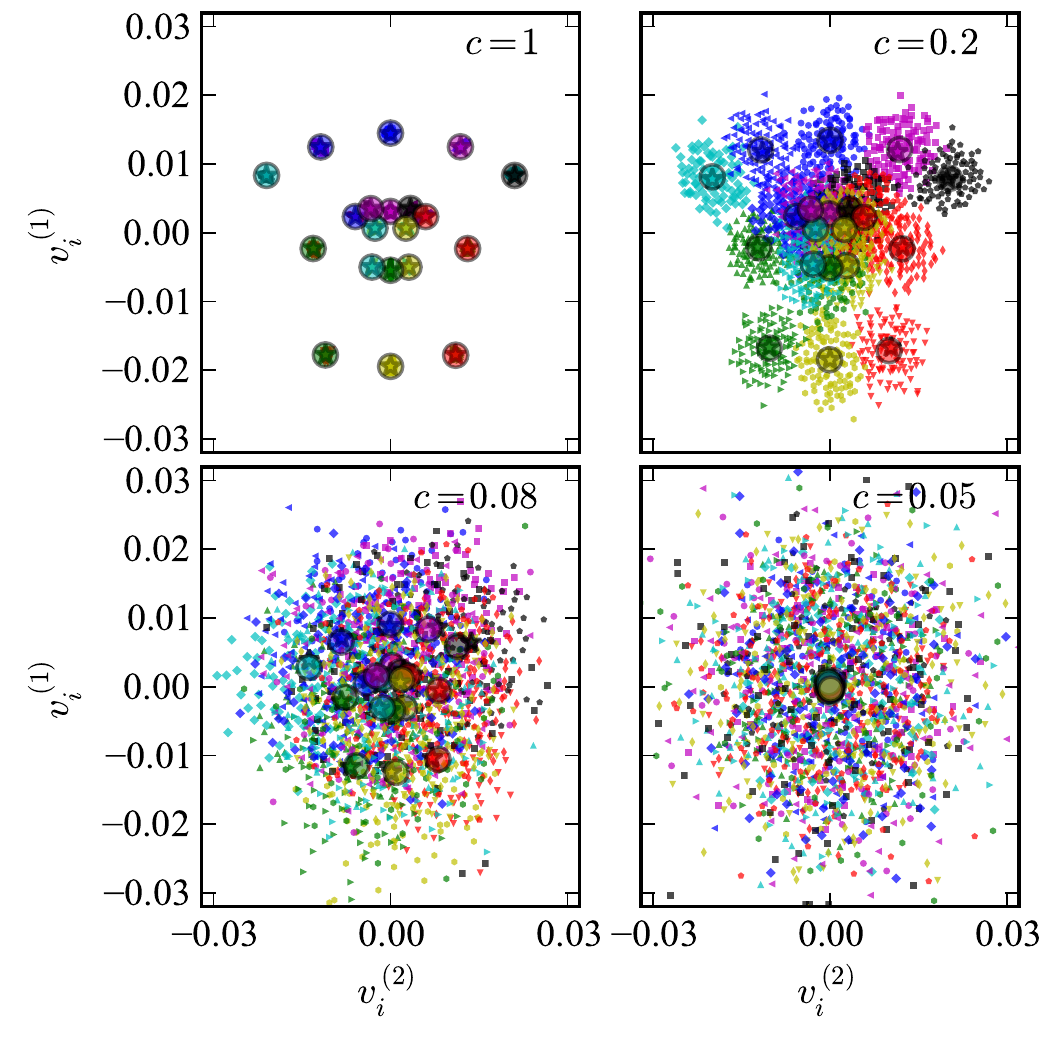}
  \end{minipage}

  \caption{\label{fig:example_profile}\emph{Left:} Extremal eigenvalues
  of $\bm{A}$ (top) and $\bm{\mathcal{L}}$ (bottom), for the block
  structure of Fig.~\ref{fig:example_adj}, as a function of the
  parameter $c$ defined in the text. The solid lines are solutions of
  Eq.~\ref{eq:detached}, and the data points are empirical values for
  $N=2\times 10^4$. The dotted vertical line marks the detachment
  transition. \emph{Right, top (bottom):} Eigenvector values for second
  and third largest (smallest) eigenvalues of $\bm{A}$
  ($\bm{\mathcal{L}}$), for different values of $c$. The circles (stars)
  correspond to the empirical (theoretical) average values for each
  block.}
\end{figure}

\emph{Eigenvectors. ---} The eigenvector equation $(\bm{\mathcal{X}} +
\bm{M}) \bm{v} = z \bm{v}$ can be rewritten as
$(z\bm{I}-\mathcal{X})^{-1} \bm{M}\bm{v} = \bm{v}$. Taking the ensemble
average, we get $\avg{(z\bm{I}-\mathcal{X})^{-1}} \bm{M}\avg{\bm{v}} =
\avg{\bm{v}}$. Since the average values of $\bm{v}$ can only depend on
the block memberships, and $\avg{(z\bm{I}-\mathcal{X})^{-1}}$ is
diagonal we get
\begin{equation}
  \bm{T}(z)\bm{M}_B\bm{N}\bm{v}_B = \bm{v}_B,
\end{equation}
where $\bm{v}_B$ contain the average values of $v$ for each block.

\begin{figure}
  \includegraphics[width=.49\columnwidth]{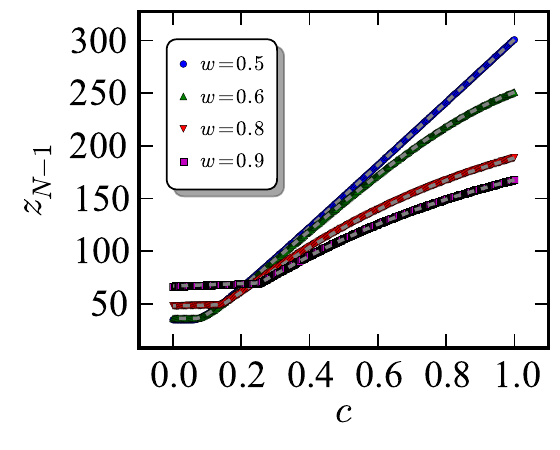}
  \includegraphics[width=.49\columnwidth]{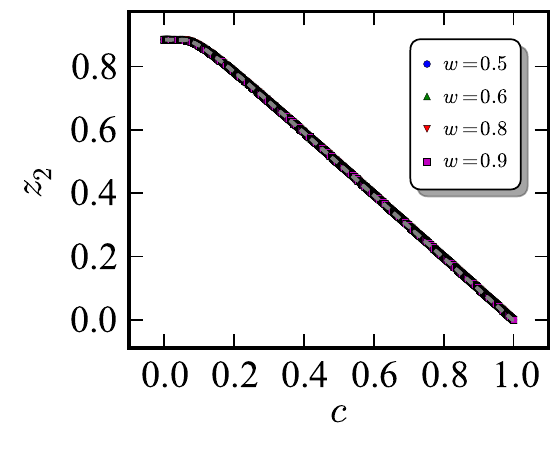}
  \includegraphics[width=.49\columnwidth]{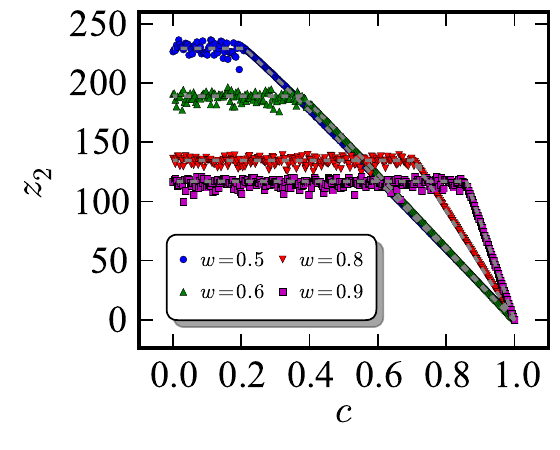}
  \includegraphics[width=.49\columnwidth]{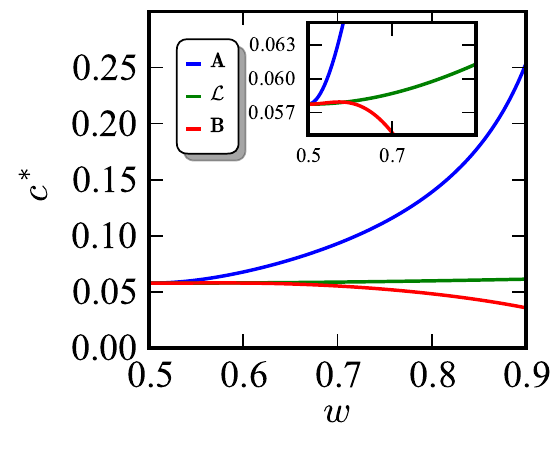}

  \caption{ \label{fig:nonhom} Top, left (right): Second largest
  (smallest) eigenvalue of $\bm{A}$ ($\bm{\mathcal{L}}$), for the
  asymmetric two-block structure described in the text. The dashed
  curves are the theoretical values, and the data points are obtained
  from network realizations with $N=2\times10^4$ and
  $\avg{k}=300$. Bottom, left: Second smallest eigenvalue of $L$. The
  dashed curves are the expected values for $N=2\times10^4$ (see
  Supplemental Material). Bottom, right: Transition point $c^*$ as a
  function of $w$ for the matrices $\bm{A}$, $\bm{\mathcal{L}}$ and the
  modularity matrix $\bm{B}$.}
\end{figure}

If the block structure is made sufficiently tenuous, all but the most
extremal detached eigenvalues will approach progressively the continuous
band. At some point, before the graph becomes fully random, they will
merge with the continuous band, and the associated eigenvectors will no
longer convey any information on the existing block structure. An
example is shown in Fig.~\ref{fig:example_profile}, which shows the full
spectrum of the block structure given by $e_{rs} = c e_{rs}^0 +
(1-c)e^0_re^0_s/2E$, with $e_{rs}^0$ being the same block structure
shown in Fig.~\ref{fig:example_adj}, and $e^0_r=\sum_se_{rs}^0$. The
parameter $c$ interpolates between a random graph ($c=0$) and the
original block structure ($c=1$), while preserving the same degree
distribution. As show in Fig.~\ref{fig:example_profile}, for a specific
value of $c = c^* > 0$ all but the most extremal eigenvalue merge with
the continuous band, and for $c < c^*$ the eigenvector values are no
longer discernibly correlated with the planted block structure. It is
important to notice that the transition point $c^*$ is different for the
matrices $\bm{A}$ and $\bm{\mathcal{L}}$, and thus the different spectra
will have different sensitivities to the planted block structure. This
can be seen in more detail by considering a simpler two-block system
with $n_1/N = w$, $n_2/N = 1-w$ and $e_{rs} = E[c\delta_{rs} +
(1-c)/2]$, which is a diagonal block structure with the parameter $c$
controlling the block segregation and $w$ the degree
asymmetry~\footnote{Note that the parameter $c$ does not change the
degree distribution.}. In Fig.~\ref{fig:nonhom} is shown the extremal
eigenvalues for the three matrices as a function of $c$, compared with
empirical values. For the normalized Laplacian matrix
$\bm{\mathcal{L}}$, the extremal eigenvalue is very insensitive to the
parameter $w$~\footnote{The curves \emph{do} change, however only very
subtly.}. The matrix $\bm{A}$ displays, on the other hand, different
transition points, depending on $w$, with larger values of $c^*$ for
larger degree asymmetries. The spectral band for the matrix $\bm{L}$ has
no well-defined edge; hence, the transition point on a finite network
will depend on the system size. The observable edge of the band is
obtained by computing the extremal statistics of $\rho(z)$ (see the
Supplemental Material), and matches well the observed values, as can be
seen in Fig.~\ref{fig:nonhom}. A comparison of the transition points can
be seen in the lower right of Fig.~\ref{fig:nonhom}, where it is also
included the values for the modularity matrix $\bm{B}=\bm{A} -
\bm{k}\bm{k}^T/2E$, where $\bm{k}$ is a vector with node degrees, often
used for community detection~\cite{newman_finding_2006}, which can also
be calculated with the presented method in an entirely analogous
fashion. Since for this specific block structure it has systematically
the lowest threshold $c^*$ among the others, this seems to corroborate
the hypothesis in Refs.~\cite{nadakuditi_graph_2012,
radicchi_detectability_2013} that $\bm{B}$ may posses optimal
characteristics in some scenarios. On the other hand, the comparatively
worst behavior of the Laplacian $\bm{L}$ raises issues with its use for
this purpose (as in e.g. Ref.~\cite{newman_community_2013}).

\emph{Homogeneous blocks.---} Further analytical progress can be made by
assuming that the blocks are homogeneous, such that the right-hand side
of Eq.~\ref{eq:t} is the same for all blocks. This means that they must
all share the same properties such as size $n_r$ and average degree
$e_r/n_r$. The solution in case $p^r_c = \delta_{d,c_r}$ (i.e. for both
$\bm{A}$ and $\bm{\mathcal{L}}$) will then be simply $t(z) = (z - d \pm
\sqrt{(d-z)^2-4a})/2a$ with $a=a_r=N\sum_s\sigma_{rs}^2/B$, which will
result in the usual semicircle distribution $\rho(z) = \sqrt{4a -
(z-d)^2}/2a\pi$ for $|z-d| < 2\sqrt{a}$; otherwise, $\rho(z) = 0$. The
detached eigenvalues will be given by the solution of $\det(\bm{I}-
t(z)N\bm{M}_B/B) = 0$. Hence there will be a one-to-one correspondence
between the nonzero eigenvalues $\lambda_i$ of $\bm{M}_B$ and the
detached eigenvalues $z_i = d + a t_i + 1/t_i$, where $t_i = B /
N\lambda_i$, as long as $|z_i-d|> 2\sqrt{a}$; otherwise, they will merge
with the continuous band. By making $|z_i-d| = 2\sqrt{a}$, one obtains
that this transition happens at $\lambda_i = \pm \sqrt{a}B/N$. Both for
$\bm{A}$ and $\bm{\mathcal{L}}$ one can see that this transition occurs
at the same point: If one writes the block matrix as $e_{rs} =
N\avg{k}m_{rs}$, such that $\sum_{rs}m_{rs} = 1$, this transition
translates to
\begin{equation}\label{eq:htrans}
  \lambda_m^2 = \frac{1}{\avg{k}B^2},
\end{equation}
where $\lambda_m$ is an eigenvalue of the $m_{rs}$ matrix. The fact that
the detachment transition is identical for both $\bm{A}$ and
$\bm{\mathcal{L}}$ is a special property of the homogeneous block
structure, and does not hold in general, as we have shown
previously~\footnote{It can also be shown that Eq.~\ref{eq:htrans} also
holds for the modularity matrix $\bm{B}$.}.

\begin{figure}
  \hspace{-1em}\includegraphics[width=.49\columnwidth]{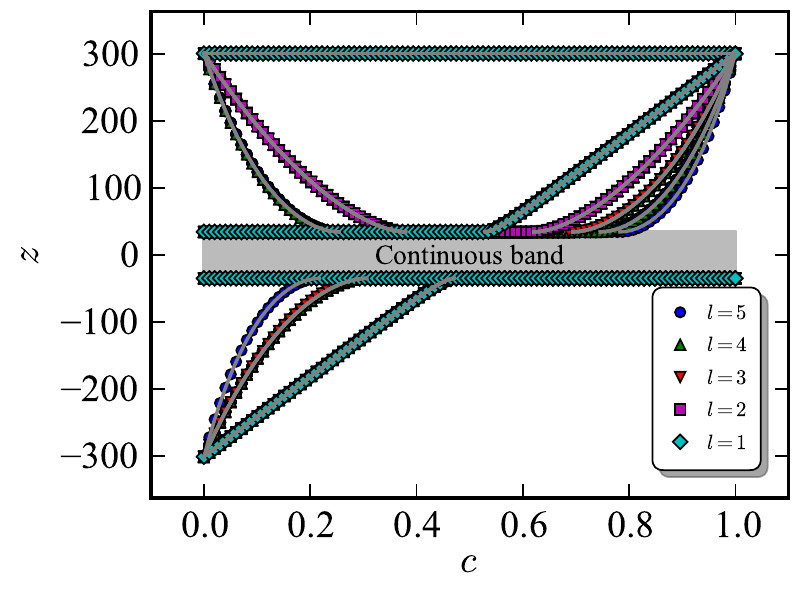}
  \includegraphics[width=.49\columnwidth]{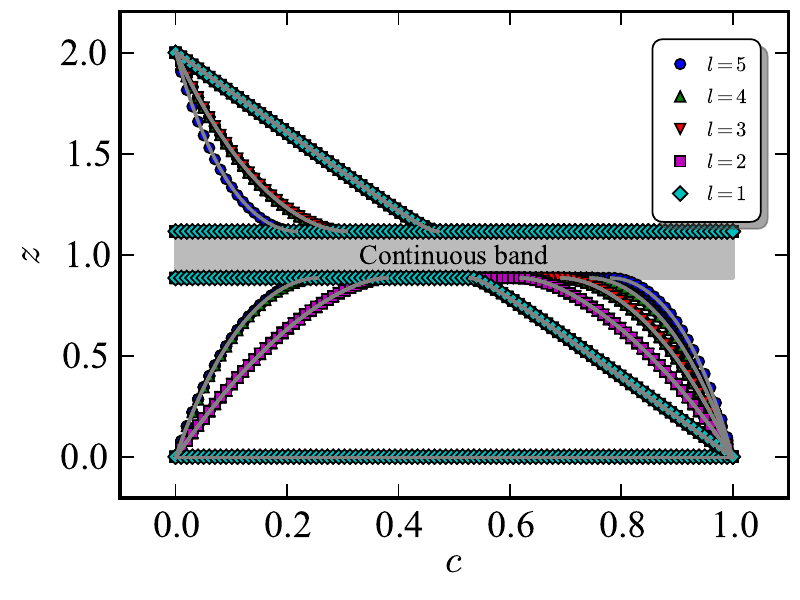}
  \begin{minipage}{.19\columnwidth}
    \centering
    \includegraphics[width=\columnwidth]{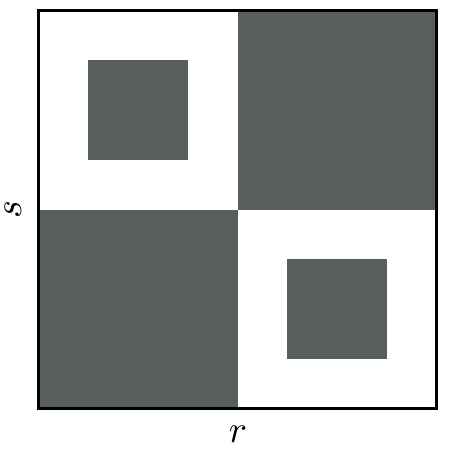}\\
    $l=1$
  \end{minipage}
  \begin{minipage}{.19\columnwidth}
    \centering
    \includegraphics[width=\columnwidth]{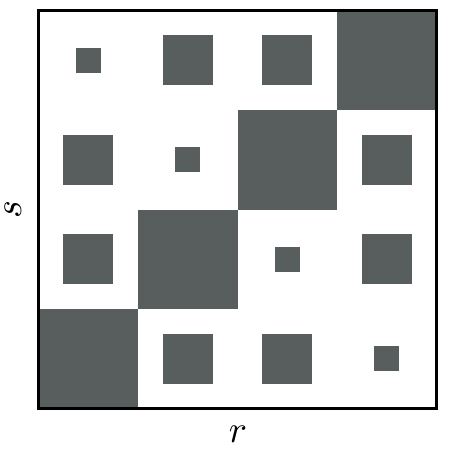}\\
    $l=2$
  \end{minipage}
  \begin{minipage}{.19\columnwidth}
    \centering
    \includegraphics[width=\columnwidth]{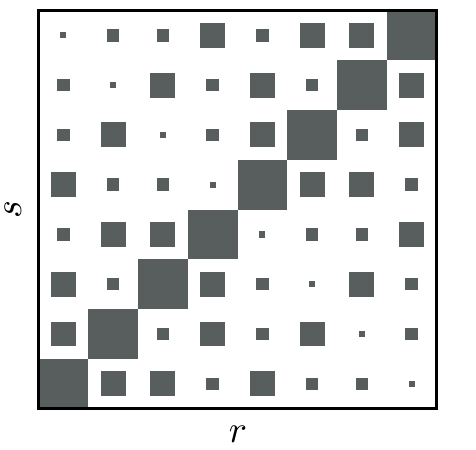}\\
    $l=3$
  \end{minipage}
  \begin{minipage}{.19\columnwidth}
    \centering
    \includegraphics[width=\columnwidth]{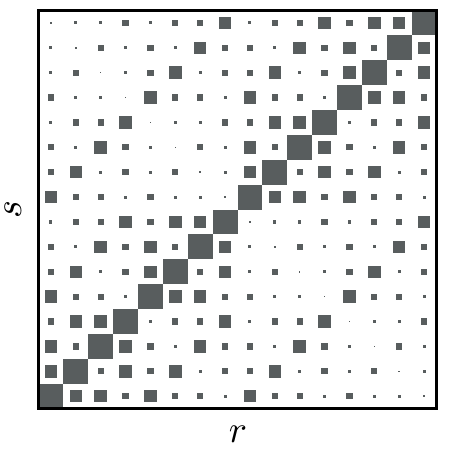}\\
    $l=4$
  \end{minipage}
  \begin{minipage}{.19\columnwidth}
    \centering
    \includegraphics[width=\columnwidth]{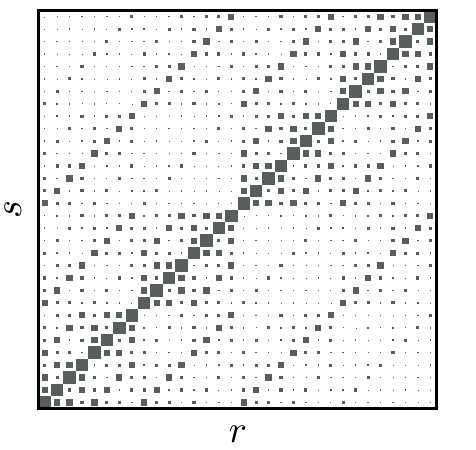}\\
    $l=5$
  \end{minipage} \caption{ \label{fig:example_nested}
    Top: Detachment transitions for the nested partition model
    described in the text with $B_1 = 2$ as a function of the mixing
    parameter $c$, and for different nesting depths $l$, for $\bm{A}$ and
    $\bm{\mathcal{L}}$. The data points correspond to network
    realizations with $N=2\times 10^4$ and $\avg{k}=300$, and the solid
    lines are theoretical values. Bottom: Example of $e_{rs}$ matrices
    with $B_1=2$ for different values of $l$.}
\end{figure}

As a concrete example of an homogeneous structure, we consider a nested
version of the usual planted partition
model~\cite{condon_algorithms_2001}, inspired by similar constructions
done in Refs.~\cite{leskovec_kronecker_2008,palla_multifractal_2010}. We
define a seed structure with $B_1$ blocks and
$[\bm{m}_1]_{rs}=\delta_{rs}c/B_1 + (1-\delta_{rs})(1-c)/B_1(B_1-1)$,
and construct a nested matrix of depth $l$ via $\bm{m}_l = \bm{m}_{l-1}
\otimes \bm{m}_{l-1}$ where $\otimes$ denotes the Kronecker product. The
eigenvalues of the matrix $\bm{m}_l$ are given by $\lambda^i_{m_l} =
((cB_1 -1)/(B_1(B_1-1)))^{l-i}/{B_1^i}$, for $i \in [0, l]$. Thus, from
Eq.~\ref{eq:htrans} one obtains a series of transitions, where a deeper
level of the nested structure ``fades away,'' and the spectrum is
indistinguishable from that of a $l-1$ structure (see
Fig.~\ref{fig:example_nested}). The transition of the shallowest level
happens at $\avg{k} = ((B-1) / (cB-1))^2$, which is the same as the
regular planted partition model~\cite{nadakuditi_graph_2012}. This
transition marks the point at which more general inference methods
should also fail to detect the imposed
partition~\cite{decelle_inference_2011}.

In summary, we presented an unified framework to obtain the full
spectrum of random networks with modular structure, in the limit of
large degrees. We showed that the detachment transition of the isolated
eigenvalues is a general feature which determines how strongly the
existing modular structure affects the different spectra. The different
matrices react differently to the imposed modular structure and have
different transition points. Only when the blocks are homogeneous do
some of these transitions collapse together. Hence, in general, the
detectability threshold of the imposed block structure may depend
strongly on the actual spectrum which is observed.

\bibliographystyle{apsrev4-1} \bibliography{bib}

\newpage\mbox{}\newpage
\onecolumngrid


\section*{Supplementary material}

# Supplemental Material: Eigenvalue spectra of modular networks

Tiago P. Peixoto*

*Institut für Theoretische Physik, Universität Bremen, Hochschulring 18, D-28359 Bremen, Germany*

## I. SYSTEM SIZE DEPENDENCE OF THE MEASURABLE SPECTRUM OF THE LAPLACIAN MATRIX $L$

The developed theoretical framework reproduces the empirical features of the Laplacian matrix $L$ in great detail, as shown in Fig. 2 of the main text, which shows that the elaborate fringe structure observed matches well the empirical observations. Such a fringed structure permeates the whole spectrum, making the definition of the edge of the continuous band quite arbitrary. However, since the amplitudes of these fringes decay when moving away from the bulk of the distribution, the probability of a given eigenvalue range may become too small to be observed in a network with a finite size. As an illustration, let us consider the probability $\rho^M(z|N)dz$ of observing a given value in the range $[z, z+dz]$ among the $M$ smallest eigenvalues which belong to the continuous band, after $N$ samples were made, which is given by the PDF,

$$\rho^M(z|N) = \binom{N}{M-1}[1-F(z)]^{N-M+1}F(z)^{M-1}\rho(z), \qquad (1)$$

where $F(z) = \int_0^z \rho(z')dz'$ is the cumulative distribution. For simplicity, let us consider a fully random $B=1$ Erdős–Rényi network as an example. In Fig. 1 is shown the distribution of the $M=100$ smallest eigenvalues for a random network of size $N = 2 \times 10^4$, together with the probability density $\rho^1(z|N)$ of the smallest eigenvalue. The latter is a fairly broad distribution, which moves very slowly to the left as the size of the network $N$ is increased. Hence the *observable* edge of the band has a relatively strong dependence on the system size, changing roughly logarithmically with $N$ (see Fig. 2).

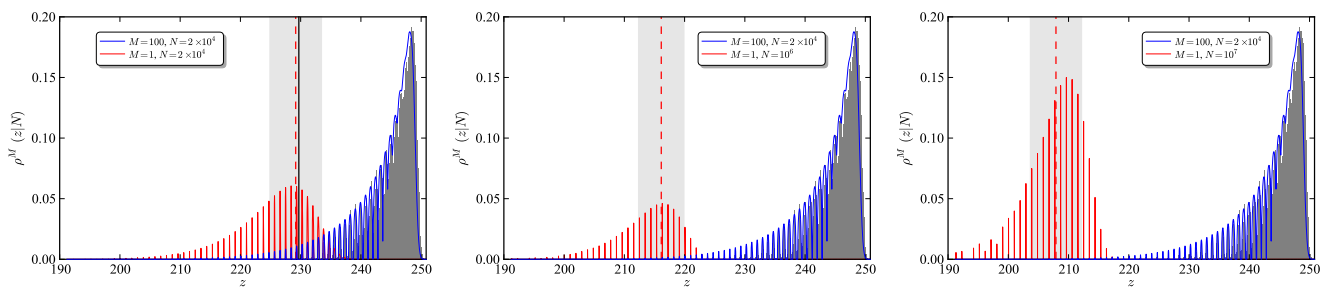

FIG. 1. PDF of the $M = 100$ smallest eigenvalues of the Laplacian matrix $L$ for an Erdős–Rényi network with $N = 2 \times 10^4$ nodes, and average degree $\langle k \rangle = 300$. The blue lines correspond to Eq. 1, and the grey histogram corresponds to 1000 independent network realizations. The red lines correspond to Eq. 1 for $M = 1$ and $N = 2 \times 10^4$ (left), $N = 10^6$ (middle) and $N = 10^7$ (right). The vertical dashed red line marks the average of the distribution $\rho^1(z|N)$, and the shaded region the standard deviation. On the leftmost plot, the vertical black line shows the empirical average for 1000 samples with $N = 2 \times 10^4$.

---

* tiago@itp.uni-bremen.de

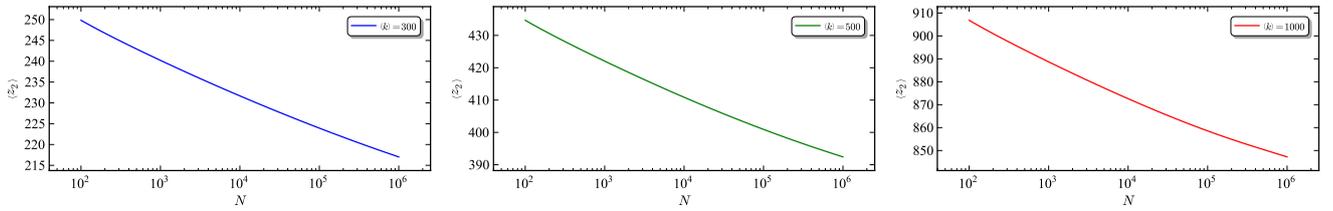

FIG. 2. Average value of the smallest nonzero eigenvalue $z_2$ of the Laplacian matrix $\boldsymbol{L}$ as a function of system size, obtained via Eq. 1, for different average degrees.

It is easy to understand such a dependence on the system size by remembering that the extremal eigenvalues of the Laplacian matrix are bounded by the maximum and minimum degrees, $k_{\max}$ and $k_{\min}$ [1, 2], as

$$z_2 \leq \frac{N}{N-1} k_{\min}, \quad z_N \geq \frac{N}{N-1} k_{\max}. \qquad (2)$$

Both for the Erdős–Rényi and the stochastic block model networks considered in the main text one should have $k_{\min} \to 0$ and $k_{\max} \to \infty$ as $N \to \infty$, so that the observable edges of the band move slowly to $z_2 \to 0$ and $z_N \to \infty$ as the system size increases, which is what is observed in Figs. 1 and 2.

---



\end{document}